\documentclass[aps,amssymb,amsmath,reprint]{revtex4-1}
\usepackage{graphicx,color}
\usepackage{epstopdf}
\usepackage{lineno}
\newcommand*\patchAmsMathEnvironmentForLineno[1]{
  \expandafter\let\csname old#1\expandafter\endcsname\csname #1\endcsname
  \expandafter\let\csname oldend#1\expandafter\endcsname\csname end#1\endcsname
  \renewenvironment{#1}
     {\linenomath\csname old#1\endcsname}
     {\csname oldend#1\endcsname\endlinenomath}}
\newcommand*\patchBothAmsMathEnvironmentsForLineno[1]{
  \patchAmsMathEnvironmentForLineno{#1}
  \patchAmsMathEnvironmentForLineno{#1*}}
\AtBeginDocument{
\patchBothAmsMathEnvironmentsForLineno{equation}
\patchBothAmsMathEnvironmentsForLineno{align}
\patchBothAmsMathEnvironmentsForLineno{flalign}
\patchBothAmsMathEnvironmentsForLineno{alignat}
\patchBothAmsMathEnvironmentsForLineno{gather}
\patchBothAmsMathEnvironmentsForLineno{multline}
}
\usepackage{booktabs}
\usepackage{comment}
\usepackage{threeparttable}
\usepackage{soul}
\usepackage{ulem}
\usepackage{here}
\usepackage{physics}
\usepackage{color}
\newif\iffigure
\figuretrue

\draft % marks overfull lines with a black rule on the right

\begin{document}

%\preprint{}

\title{
Different spin relaxation property observed in linearly and circularly polarized laser induced terahertz emission from Bi/Co bilayer}


\author{Kazuaki Ishibashi$^{1,2}$}
\email{kazuaki.ishibashi.p1@dc.tohoku.ac.jp}
\affiliation{$^1$Department of Applied Physics, Graduate School of Engineering, Tohoku University, Sendai 980-8579, Japan}
\affiliation{$^2$WPI Advanced Institute for Materials Research (AIMR), Tohoku University, 2-1-1, Katahira, Sendai 980-8577, Japan}
\affiliation{$^3$Frontier Research Institute for Interdisciplinary Sciences (FRIS), Tohoku University, Sendai 980-8578, Japan}
\affiliation{$^4$Center for Science and Innovation in Spintronics (CSIS), Core Research Cluster (CRC), Tohoku University, Sendai 980-8577, Japan}

\author{Satoshi Iihama$^{3,2}$}
\email{satoshi.iihama.d6@tohoku.ac.jp}
\affiliation{$^1$Department of Applied Physics, Graduate School of Engineering, Tohoku University, Sendai 980-8579, Japan}
\affiliation{$^2$WPI Advanced Institute for Materials Research (AIMR), Tohoku University, 2-1-1, Katahira, Sendai 980-8577, Japan}
\affiliation{$^3$Frontier Research Institute for Interdisciplinary Sciences (FRIS), Tohoku University, Sendai 980-8578, Japan}
\affiliation{$^4$Center for Science and Innovation in Spintronics (CSIS), Core Research Cluster (CRC), Tohoku University, Sendai 980-8577, Japan}
%\affiliation{WPI Advanced Institute for Materials Research (AIMR), Tohoku University, 2-1-1, Katahira, Sendai 980-8577, Japan}

\author{Shigemi Mizukami$^{2,4}$}
\affiliation{$^1$Department of Applied Physics, Graduate School of Engineering, Tohoku University, Sendai 980-8579, Japan}
\affiliation{$^2$WPI Advanced Institute for Materials Research (AIMR), Tohoku University, 2-1-1, Katahira, Sendai 980-8577, Japan}
\affiliation{$^3$Frontier Research Institute for Interdisciplinary Sciences (FRIS), Tohoku University, Sendai 980-8578, Japan}
\affiliation{$^4$Center for Science and Innovation in Spintronics (CSIS), Core Research Cluster (CRC), Tohoku University, Sendai 980-8577, Japan}

%\listoffigures


\date{\today}

\begin{abstract}
Recently, helicity-dependent photocurrent was reported in Bi single thin films. 
It is proposed that the origin of this photocurrent is the combination of photo-spin conversion and spin-charge conversion effects in Bi and efficient spin conversion in Bi is expected. 
In this study, we measured two types of terahertz (THz) emissions from Bi/Co bilayer films induced by spin current generation using laser-induced demagnetization of the Co layer and photo-spin conversion effect in the Bi layer to investigate the spin current induced by the two mechanisms simultaneously. 
We clearly observed different Bi thickness dependence of peak intensity and that of bandwidth for THz spin current in two experiments, {\it i.e.}, spin current induced by demagnetization of Co and that by photo-spin conversion in Bi.
The different Bi thickness dependence of spin current intensity and bandwidth in two experiments is caused by different spin relaxation properties of optically excited spin currents in Bi layers.

\end{abstract}
\maketitle

\section{Introduction}
Conversion between electron spin and physical quantities, such as charge, light, heat, and phonon, is one of the fundamental principles that enable the generation and detection of the spin current\cite{Sinova_2015, Choi_2017, Uchida_2010, Kawada_2021}. 
To enhance the conversion efficiency for future spintronic devices, numerous studies have explored various materials, such as topological materials\cite{Kondou_2016,Wu_2021} and heavy metals (e.g., Pt, W, Ta, and Bi\cite{Tanaka_2008,Ando_2008, Pai_2012, Liu_2012, Hou_2012, Matsushima_2020}).
In particular, Bi is the basis of several topological materials, such as Bi$_{0.9}$Sb$_{0.1}$, PtBi$_{2}$, and Bi$_{x}$Se$_{1-x}$\cite{Khang_2018, Gibson_2015, Dc_2018}. Accordingly, Bi-based alloys are among the candidate materials with good spin current generation characteristics owing to their large spin orbit coupling and unique band structure. 
Thus, it is important to investigate the phenomena occurring in Bi for obtaining a basic understanding of the phenomena occurring in Bi-based alloys.
In addition, Bi itself is expected to exhibit an efficient spin conversion effect and other interesting phenomena\cite{Fuseya2015}.

Recently, helicity-dependent (HD) photocurrent in Bi single thin film or Bi/Cu(or Ag) bilayer films was observed via pulse laser-induced terahertz (THz) emission\cite{Hirai_2020} and transport measurement using a continuous wave laser\cite{Hirose_2021}. 
The proposed mechanism for photocurrent in Bi is photo-induced inverse spin-Hall effect\cite{Ando_2010}. 
A circularly polarized laser induces electron spin in Bi depending on optical helicity via conversion from photon spin angular momentum (SAM) to electron-SAM, which is called the photo-spin conversion effect. Subsequently, the flow of electron-SAM is converted to charge current through the inverse spin-Hall effect. 
Although photon-SAM driven torques were observed in heavy metal/ferromagnet bilayer films via time-resolved magneto-optical Kerr effect measurement\cite{Choi_2017, Choi_2020, Iihama_2021, Iihama_2022}, the abovementioned HD photocurrent in a single thin film has been mostly reported in Bi-related materials\cite{Hirai_2020, Hirose_2021, Kawaguchi_arxiv_2020}. Therefore, the efficient photo-spin conversion effect in Bi is expected, which most likely originates from the band structure inherent to semi-metallic Bi. 
However, the details of photo-spin conversion in Bi have not been clarified yet because the photo-spin conversion and spin-charge conversion effects are observed simultaneously in the photocurrent measurement, thus, one cannot distinguish between two spin-related conversion effects.

In this study, to disentangle the two processes, namely the photo-spin conversion and spin-charge conversion effects, and gain insight into the underlying physics, we measured the THz emission induced by spin current generation using photo-spin conversion and laser-induced demagnetization simultaneously in Bi/Co bilayer films. 

The THz emission experiment conducted with structures widely used in spintronics THz emitters\cite{Kampfrath_2013, Seifert_2016, Seifert_2018}, {\it e.g.}, ferromagnet/Bi bilayers, remains scarce.
When a femtosecond laser pulse is irradiated on the ferromagnet/Bi bilayer, spin current can be generated by the ultrafast demagnetization of the ferromagnetic layer due to the conservation of angular momentum\cite{Choi_2014, Lichtenberg_2022, Rouzegar_2022}. 
This transient spin current is converted into charge current owing to spin-charge conversion effect in the Bi layer, and then THz wave is emitted.
Thus, we can simultaneously investigate the difference in spin current via photo-spin conversion and laser-induced demagnetization using this structure.

\section{Experiment}

\begin{figure*}[ht]
\begin{center}
\includegraphics[width=0.7\textwidth,keepaspectratio,clip]{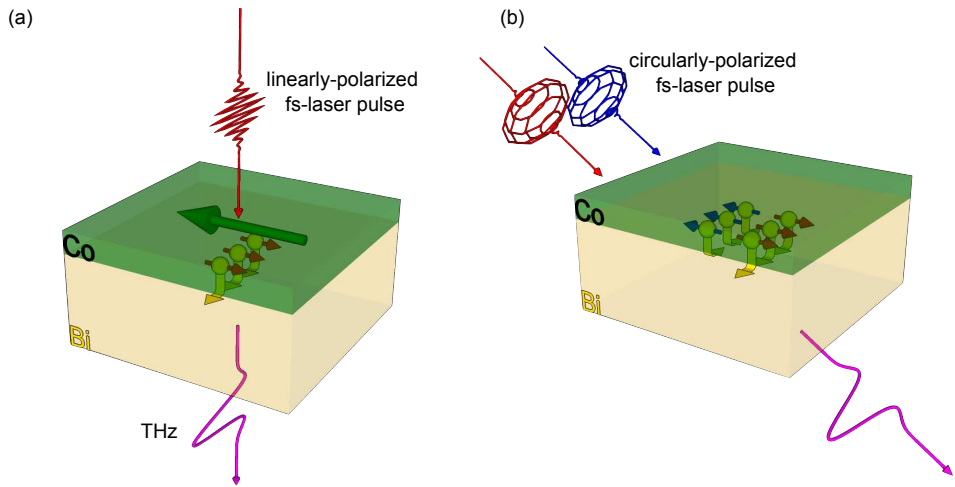}
\end{center}
\caption{ Schematic illustration of (a) linearly and (b) circularly polarized laser-induced terahertz emission from Bi/Co bilayer. }
\label{f1}
\end{figure*}

The samples were prepared using DC/RF magnetron sputtering. 
The stacking structure of samples was Glass sub./Bi($d_{\rm{Bi}}$)/Co(5)/MgO(2)/Ta(2) (thickness is in nm). 
The thicknesses of the Bi layers $d_{\rm{Bi}}$ were varied from 10 to 120 nm. 
The Co layer generates the spin current from laser-induced demagnetization. 
The Bi layers generate spin current via the photo-spin conversion effect and convert the spin current into charge current via the spin-charge conversion effect. 
The MgO and Ta layers are capping layers to prevent oxidization. 
The Bi film in all samples was polycrystalline with the (003) and (012) preferred orientations indexed using hexagonal notation. 
The saturation magnetization of Co was almost constant with respect to Bi thickness (see Appendix for details on sample information).

The laser pulse-induced THz emission from Bi/Co films was measured using THz time domain spectroscopy (THz-TDS)\cite{Sasaki_2017, Sasaki_2019, Idzuchi_2021}. 
Laser pulses are generated by a Ti:Sapphire femtosecond laser with a wavelength of 800 nm, pulse duration of 160 fs, and repetition rate of 1 kHz. 
Pump laser pulses are modulated by a mechanical chopper at a frequency of 360 Hz. 
A quarter wave plate (QWP) is placed in front of samples to control pump laser polarization. 
The pump laser was focused on the film with a fluence of 0.62 mJ/cm$^{2}$. 
The polarization of the THz wave emitted from the sample surface was analyzed with two wire grids\cite{Sasaki_2020_APL, Ogasawara_2020}. 
We measured the emitted THz wave using the electro-optic (EO) sampling method\cite{Gallot_1999} with a 1-mm-thick ZnTe(110) crystal. 
All measurements were taken at room temperature.

\begin{figure*}[ht]
\begin{center}
\includegraphics[width=0.7\textwidth,keepaspectratio,clip]{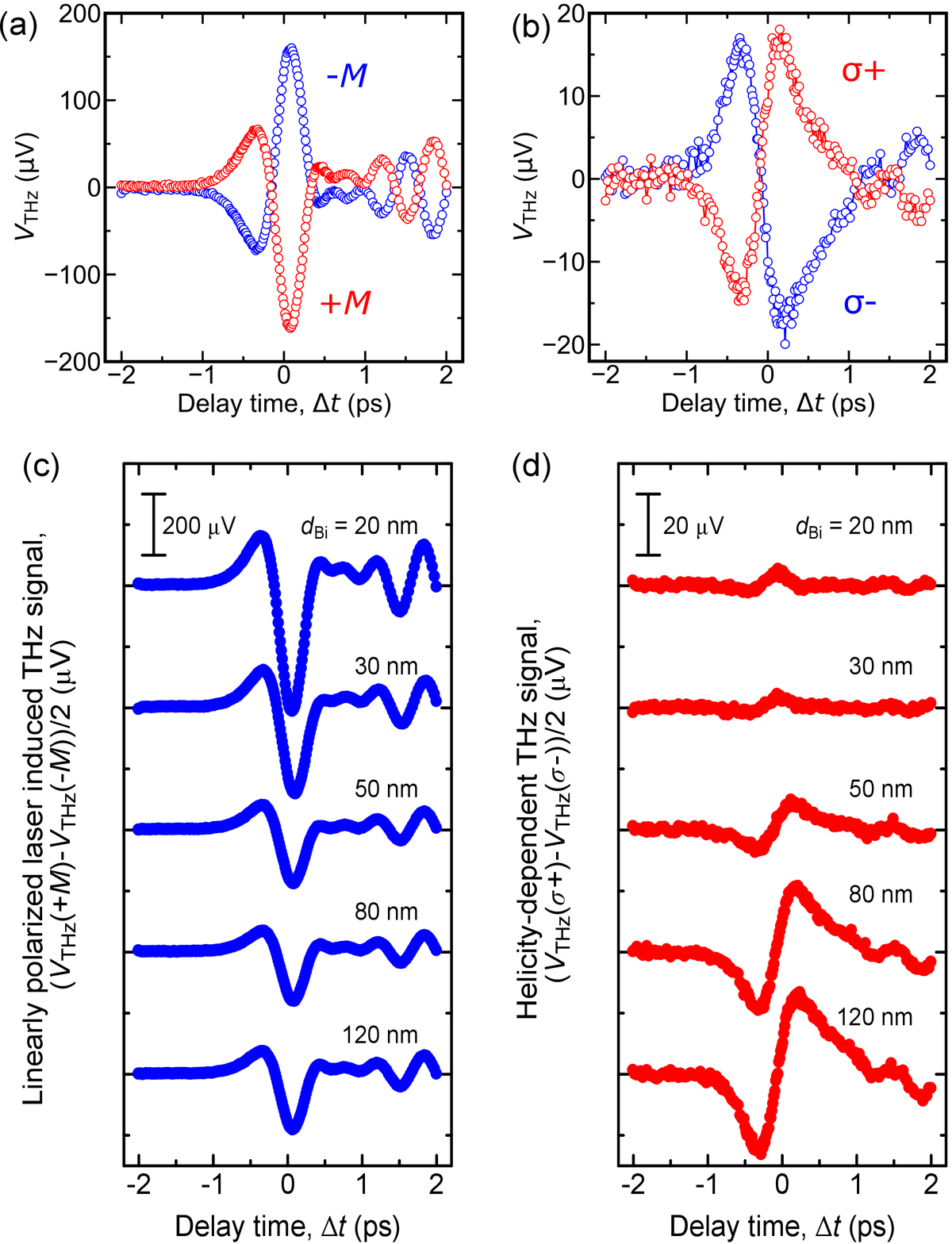}
\end{center}
\caption{ (a) Linearly polarized laser-induced terahertz signal with different polarity of Co magnetization where laser is irradiated with normal incidence. (b) Circularly polarized laser-induced terahertz signal with different helicity of laser pulse where laser is irradiated with 45 deg. incident angle. Bi thickness $d_{\rm Bi}$ dependence of (c) linearly polarized laser-induced terahertz signal and (d) HD circularly polarized laser-induced terahertz signal where difference between signal obtained with left- and right-circularly polarized light is taken. }
\label{f2}
\end{figure*}

\section{Experimental results and discussion}
 \subsection{Laser-induced THz emission and Bi thickness dependence} 

We measured two kinds of THz waves emitted from Bi/Co bilayer films induced by linearly polarized and circularly polarized lasers as shown in Figs. 1(a) and 1(b), respectively. 
The difference between the two measurements is the source of spin current. 
The first spin current source is laser-induced demagnetization. 
When a linearly polarized laser is irradiated on the sample, ultrafast spin current is generated from the laser-induced demagnetization of Co, where the polarization of spin is parallel to the magnetization direction of Co. 
This spin current flows into an adjacent Bi layer and is converted to charge current through the inverse spin-Hall effect in Bi, which causes THz emission from the film surface [Fig. 1(a)]. 
The second source is photo-spin injection via the photo-spin conversion effect in Bi. 
When a circularly polarized laser is irradiated on the sample with an oblique incidence, in-plane electron spin is injected in Bi depending on the optical helicity via the photo-spin conversion effect. 
The incident angle was fixed at 45$^{\circ }$ in our measurement. 
Spin current is caused by the gradient of induced spin because of the finite penetration depth of the laser. 
This spin current is converted into a charge current and then the THz wave is emitted [Fig. 1(b)]. 
In contrast to the linearly polarized laser-induced THz emission, circularly polarized laser-induced THz emission is observed in Bi single thin film\cite{Hirai_2020}. 
We conducted THz experiments using spin currents generated by two different mechanisms shown in Figs. 1(a) and 1(b), where the THz wave polarization induced by demagnetization is orthogonal to that induced by photo-spin injection to distinguish the two contributions via THz polarization analysis with two wire grids. 
Note that we measured photo-spin conversion induced terahertz emission with no net magnetization of Co, {\it i.e.}, samples in a virgin state were measured for THz emission induced by photo-spin conversion in Bi.
This no net magnetization was confirmed by measuring orthogonal component of THz signal.
HD-THz caused by the inverse Faraday effect and the inverse spin-orbit torque observed in the previous study\cite{Huisman2016a} can be ruled out because of no net magnetization of Co.

A typical THz signal $V_{\rm{THz}}$ induced by linearly polarized laser with two opposite sample magnetization $\pm M$ orientations is shown in Fig. 2(a). 
The THz signal is inverted when the magnetization is reversed, which is consistent with THz signals emitted from magnetic heterostructures\cite{Kampfrath_2013,Seifert_2018}. 
It was found that there is a contribution from the ordinary Nernst effect in Bi\cite{Yue_2018} where the THz signal is linearly changed based on the external magnetic field strength.
To remove the contribution of the ordinary Nernst effect, first an external magnetic field was applied to saturate the magnetization of Co and then the THz signal was measured without an external magnetic field. 
Fig. 2(b) shows typical circularly polarized laser pulse-induced THz signals $V_{\rm{THz}}$ with different optical helicities $\sigma \pm $. 
The red and blue circles represent the THz signals induced by a right- and left-circularly polarized laser, respectively. 
The sign of the THz signal is reversed when the helicity of the circularly polarized laser pulse is changed, indicating the HD-THz signal. 
To focus on the HD and magnetization direction dependent contributions, we considered antisymmetric signals with respect to magnetization and optical helicity, {\it i.e.}, $(V_{\rm{THz}} (+M) - V_{\rm{THz}} (-M))/2$ and  $(V_{\rm{THz}} (\sigma +) - V_{\rm{THz}} (\sigma -))/2$.

\begin{figure*}[ht]
\begin{center}
\includegraphics[width=0.7\textwidth,keepaspectratio,clip]{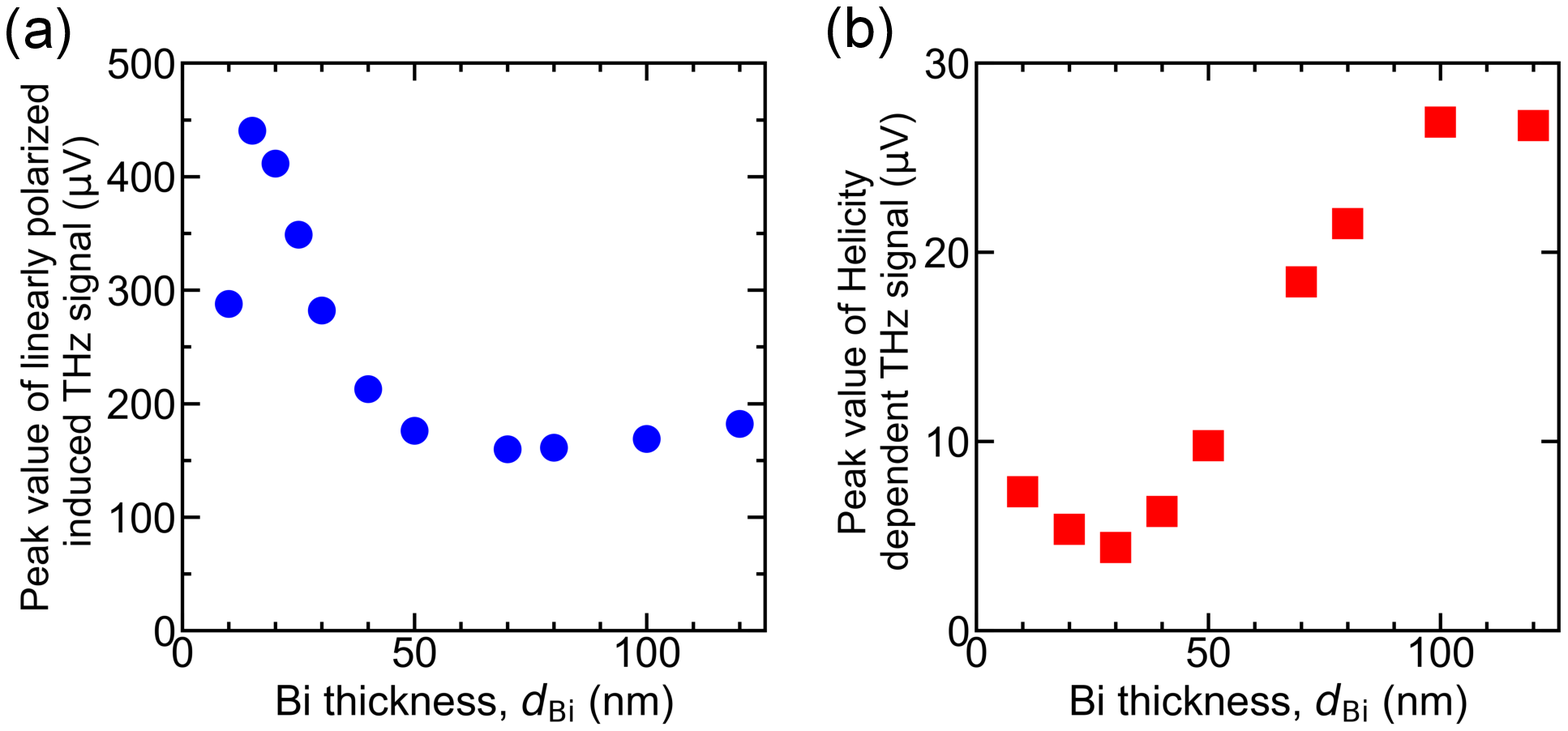}
\end{center}
\caption{ Peak value of (a) linearly polarized laser-induced terahertz signal and (b) HD terahertz signal plotted as a function of Bi thickness $d_{\rm Bi}$.}
\label{f3}
\end{figure*}

\begin{figure*}[ht]
\begin{center}
\includegraphics[width=0.7\textwidth,keepaspectratio,clip]{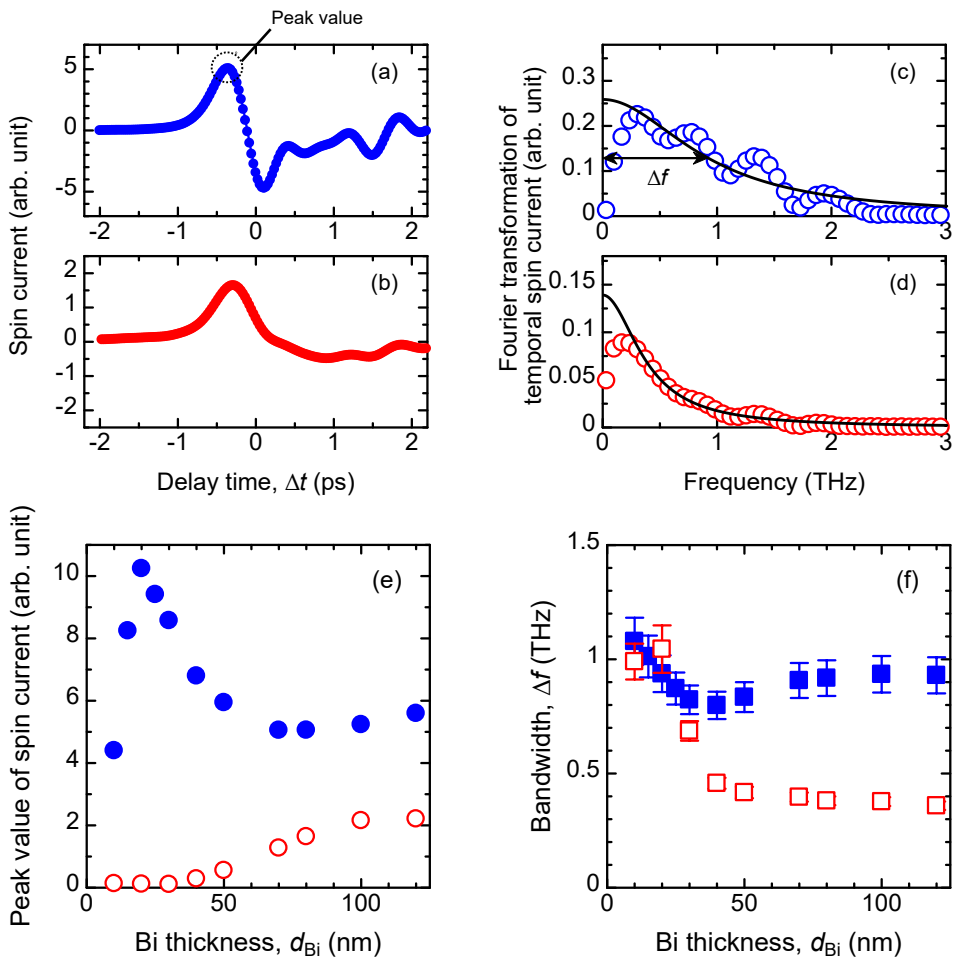}
\end{center}
\caption{ Temporal spin current signal obtained from (a) linearly polarized laser-induced terahertz signal and (b) circularly polarized laser-induced terahertz signal for Bi(80) /Co(5) bilayer. (c) Fourier transformation spectrum for spin current induced by (c) linearly polarized laser and (d) circularly polarized laser, corresponding to (a) and (b). Solid curves represent Lorentzian fitting to obtain bandwidth of spectra. (e) Peak value of spin current plotted as a function of Bi thickness $d_{\rm Bi}$. (f) Bandwidth of Fourier transformation spectrum $\Delta f$ plotted as a function of $d_{\rm Bi}$. }
\label{f4}
\end{figure*}

\begin{figure*}[ht]
\begin{center}
\includegraphics[width=0.7\textwidth,keepaspectratio,clip]{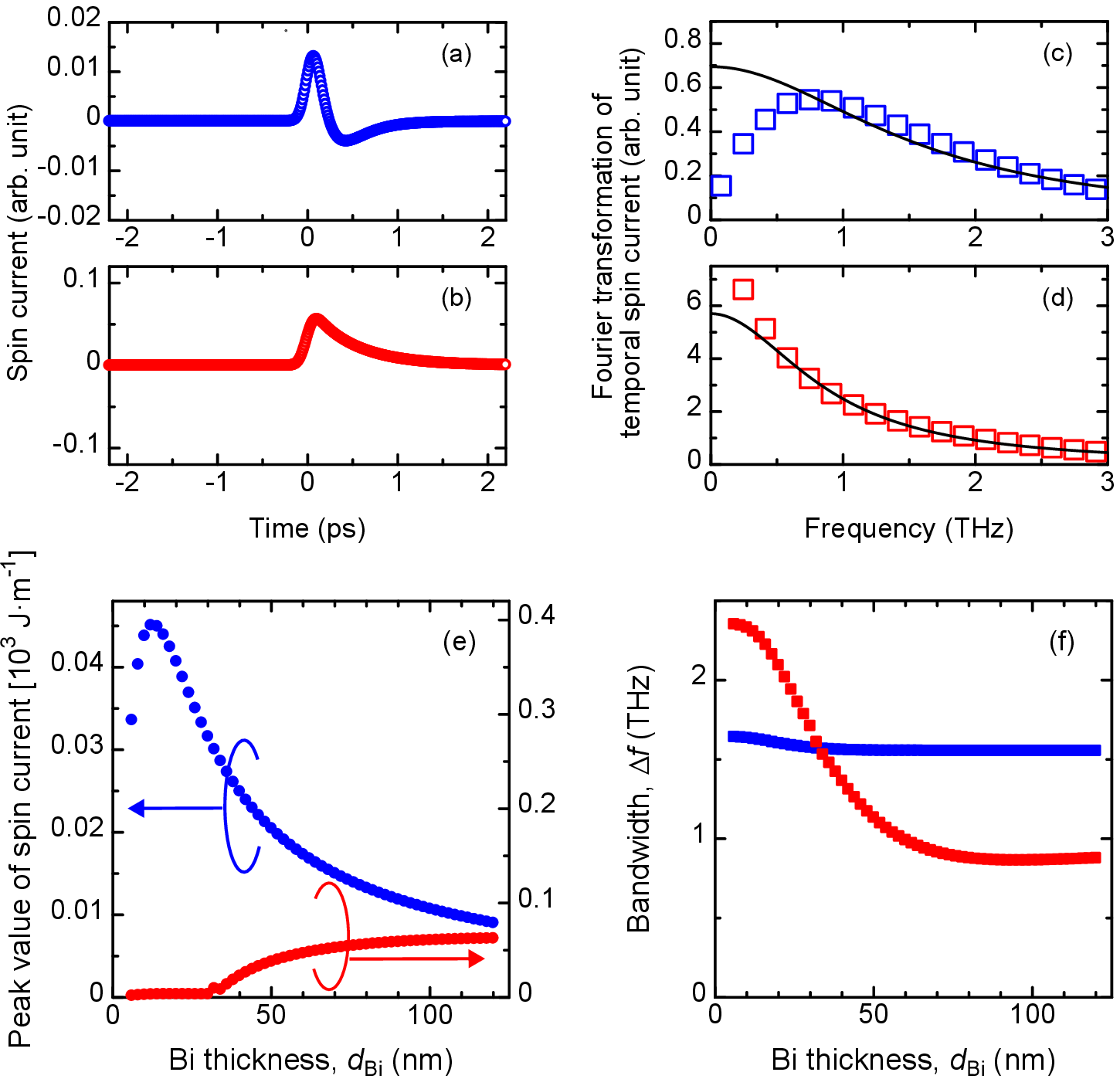}
\end{center}
\caption{ Temporal spin current induced by (a) demagnetization of Co layer and (b) photo-spin injection into Bi layer, calculated by spin-diffusion equation. (c), (d) Fourier transformation spectrum for spin current corresponding to (a), (b). (e) Peak values of spin current obtained by spin-diffusion simulation plotted as a function of Bi thickness $d_{\rm Bi}$. (f) Bandwidth of Fourier transformation spectrum $\Delta f$ obtained by spin-diffusion simulation plotted as a function of $d_{\rm Bi}$. }
\label{f5}
\end{figure*}

Figs. 2(c) and 2(d) show the Bi thickness $d_{\rm{Bi}}$ dependence of the linearly polarized laser-induced THz signal and HD circularly polarized laser-induced THz signal, respectively. 
As the Bi thickness $d_{\rm{Bi}}$ increased from 20 nm to 120 nm, the amplitude of the THz signal induced by the linearly polarized laser decreased. 
In contrast to linearly polarized laser-induced THz, the amplitude of the HD-THz signal induced by the circularly polarized laser increased. 
Figs. 3(a) and 3(b) exhibit the $d_{\rm{Bi}}$ dependence of peak value of the linearly polarized laser-induced THz signal and HD circularly polarized laser-induced THz signal, respectively. 
Different $d_{\rm{Bi}}$ dependences were clearly observed in the two experiments. The trends observed in the two experiments, shown in Figs. 3(a) and 3(b), were consistent with those observed in previous studies\cite{Seifert_2018,Hirai_2020}.
Moreover, the waveform of THz was different, {\it i.e.} its shape is broader for the HD-THz signal compared with the linearly polarized laser-induced THz signal. 
Those differences can be caused by different temporal dynamics of the laser-induced spin current. 
To obtain the temporal spin current, we analyzed THz signals as described below.

\subsection{Spin current analyzed from experiment}
THz signals measured via the EO sampling method can be converted into THz electric field in the sample and temporal spin current. 
In the frequency domain, the relation between the measured THz signal $V_{\rm{THz}}$ and THz electric field $E$ is stated as\cite{Seifert_2016}
\begin{equation}
V_{\rm{THz}}(\omega) \propto H(\omega)E(\omega), \label{eq_VHE}
\end{equation}
where $H(\omega)$ is the response function of the THz detection process. 
The response function comprises the propagation of the THz pulse from the sample to the detector and the detector response function for the ZnTe (110) crystal. 
The detection of the THz electric field is assumed to be far-field, {\it i.e.}, the response function for propagation is proportional to the frequency.
The detector response function for the ZnTe crystal was calculated with parameters taken from Ref. \cite{ZnTe_param}. 
The THz electric field generated from the laser-induced spin current is described as\cite{Seifert_2016},
\begin{equation}
E(\omega) = eZ(\omega)\theta_{\rm{SH}}\int_{0}^{d_{\rm{Bi}}} J_{\rm{s}}(z, \omega) dz,\label{eq_EJ}
\end{equation}
where $\theta_{\rm{SH}}$ and $e$ are the spin-Hall angle and electron charge, respectively. 
In addition, $J_{\rm{s}}(\omega)$ represents spin current in the frequency domain. The electromagnetic impedance $Z(\omega)$ is expressed as
\begin{equation}
Z(\omega)= \frac{Z_{0}}{n_{0}+n_{\rm{sub}}+Z_{0}\sigma_{\rm{tot}}(\omega)d_{\rm{tot}}},\label{eq_Z}
\end{equation}
where $Z_{0}, n_{0}, n_{\rm{sub}}$, and $d_{\rm{tot}}$ are the vacuum impedance, refractive index of air, refractive index of substrate, and total film thickness, respectively. 
Here, $n_0$ = 1.0 and $n_{\rm sub}$ = 1.5.
$\sigma_{\rm{tot}}(\omega)$ is the total conductivity, which was assumed to be constant in the THz range\cite{Hirai_2020, Sasaki_2020}.
$\sigma _{\rm tot}$ values were measured using the four-point probe method (see Appendix 2 for electrical conductivity of samples). 

Using Eq. (\ref{eq_VHE}) $\sim$ (\ref{eq_Z}) and the measured THz signals, the spin current was obtained with a cutoff frequency of 3 THz. 
Figs. 4(a) and 4(b) show temporal spin current $\int_{0}^{d_{\rm{Bi}}} J_{\rm{s}}(z, t) dz$ obtained from the linearly and circularly polarized laser-induced THz signal for the Bi(80)/Co(5) bilayer film, respectively. 
Figs. 4(c) and 4(d) show the Fourier transformation spectrum $\int_{0}^{d_{\rm{Bi}}} J_{\rm{s}}(z, \omega) dz$ for the spin currents depicted in Figs. 4(a) and 4(b), respectively. 
The solid curve denotes the result fitted with Lorentzian; subsequently, the bandwidth $\Delta f$ of spectrum for the spin current is evaluated. 
The peak values of the spin current and bandwidth $\Delta f$ plotted as a function of Bi thickness $d_{\rm Bi}$ are shown in Figs. 4(e) and 4(f), respectively. 
Blue solid symbols and red open symbols represent the values obtained from linearly and circularly polarized laser-induced spin currents, respectively. 
The intensity of the spin current induced by the linearly polarized laser peaked at approximately $d_{\rm{Bi}} = 20$ nm [solid circles in Fig. 4(e)], whereas the amplitude of the spin current induced by the circularly polarized laser peaked at approximately $d_{\rm{Bi}} = 100$ nm [open circles in Fig. 4(e)]. 
The $\Delta f$ of the spin current induced by the linearly polarized laser was almost constant with respect to Bi thickness $d_{\rm{Bi}}$ [solid squares in Fig. 4(f)]. 
On the other hand, the $\Delta f$ of the spin current induced by the circularly polarized laser exhibited a decreasing trend as $d_{\rm{Bi}}$ increased and then saturated to a constant [open squares in Fig. 4(f)]. 
In addition, the $\Delta f$ of the spin current induced by the circularly polarized laser was one half of that induced by the linearly polarized laser when Bi thickness is sufficiently thick. 
The differences can be caused by the difference in the source of spin current and spin-transport discussed in the next section.

\subsection{Theoretical analysis of spin current}

To explain the $d_{\rm Bi}$ dependence of laser-induced spin current, we performed simulation of the spin-diffusion equation.
Superdiffusive transport as well as Boltzmann transport approaches more rigorously depict interacting electron kinetics with various energy scales far from thermal equilibrium state\cite{Battiato_2010, Nenno_2018}. However, these models are quite complex since one needs to consider electron energy, momentum and scattering properly.
In the diffusive approach, \color{black}backflow spin current cannot be reproduced and \color{black}it is implicitly assumed that spin angular momentum are moved based on the continuity euation and there are many electron scattering during spin transport.The spin-transport in the spin-diffusion equation can be simply described by the two parameters how fast spin angular momentum is diffused and lost.
The spin-diffusion equation is described as follows\cite{Stiles_2004}:
\begin{equation}
\pdv{s(z,t)}{t} = D\pdv[2]{s(z,t)}{z} - \frac{s(z,t)}{\tau_{\rm{s}}} +Q_{\rm{s}}(z,t),\label{eq_spindiffusion}
\end{equation}
where $s$, $D$, and $\tau_{\rm{s}}$ denote the electron-SAM density, diffusion constant, and spin relaxation time, respectively. 
Here, $Q_{\rm{s}}$ corresponds to the source term for the spin current, and we assumed two different spin current sources in two experiments as described below. 
First, demagnetization is considered as a spin source at the interface between the Co and Bi layers $Q_{\rm{s}}^{\rm{d}}$, which can be expressed as follows:
\begin{equation}
Q_{\rm{s}}^{\rm{d}}(z,t)  = \left\{ 
\begin{array}{cc}
-\frac{d_{\rm{Co}}}{\gamma}\frac{d}{dt}(\Delta M_{\rm{s}}(t; d_{\rm Bi})) &\hspace{0.5cm} {\rm if} \hspace{0.2cm} z=d_{\rm Bi}, \\
0 & {\rm else}.
\end{array}
\right.,\label{eq_Q_d}
\end{equation}
where $\gamma$ and $\Delta M_{\rm{s}}$ denote the gyromagnetic ratio and temporal dynamics of demagnetization, respectively. 
$\Delta M_{\rm{s}}$ was evaluated by using the time-resolved magneto-optical Kerr effect (TRMOKE) measurement\cite{Iihama_2015} with a constant pump fluence of 0.62 mJ/cm$^{2}$. 
The laser-induced spin current has been considered to be inversely proportional to the total layer thicknesses for spintronic THz emitters described in a previous study\cite{Seifert_2016}.
In fact, laser-induced demagnetization decreased with increasing $d_{\rm Bi}$ values (see Appendix 4 for $d_{\rm Bi}$ dependence of demagnetization), which is consistent with the assumption that the absorbed fluence per unit thickness decreases with an increase in the metallic layer thickness. 
Therefore, we considered $d_{\rm{Bi}}$ dependence of demagnetization $\Delta M_{\rm{s}}(t;d_{\rm Bi})$, which is proportional to $(d_{\rm Bi}+d_{\rm Co})^{-1}$, {\it i.e.}, $\Delta M_{\rm{s}}(t;d_{\rm{Bi}}) = \Delta M_{\rm{s}}(t;d_{\rm{Bi}} = 15) \cdot (15+d_{\rm Co}) / (d_{\rm{Bi}} + d_{\rm{Co}}) $ using the demagnetization dynamics for a 15-nm-thick Bi sample, $\Delta M_{\rm{s}}(t;d_{\rm{Bi}} = 15)$.

On the other hand, photo-spin injection is considered as a spin current source for circularly polarized laser-induced THz signals. 
Photo-spin conversion-induced spin density $Q_{\rm{s}}^{\rm{p}}$ can be considered as the conversion between absorbed photon-SAM and electron-SAM, which is described by the following equation\cite{Iihama_2022}:
\begin{equation}
Q_{\rm{s}}^{\rm{p}}(z,t) = \frac{\eta a(z)F_{\rm{p}}}{\omega _{\rm l}} \sin\theta_{\rm{inc}}G(t),\label{eq_Q_p}
\end{equation}
where $F_{\rm{p}}, \omega_{\rm l}, \theta_{\rm{inc}}$, and $G(t)$ denote the fluence of pump laser, laser angular frequency, incident angle of laser and temporal profile of the Gaussian laser pulse, respectively. 
The laser absorption profile inside the Bi layer $a(z)$ is calculated using the transfer matrix method\cite{Zak_1990} (see Appendix for refractive index and light absorption). 
Here, we assumed that photon-SAM is entirely converted into electron-SAM, {\it i.e.}, $\eta $ = 1.
The obtained $s(z,t)$ can be converted into spin current via the following relation:
\begin{equation} 
J_{\rm{s}} (z,t)= D\pdv{s(z,t)}{z}, \label{eq_spin current}
\end{equation}
Using Eq. (\ref{eq_spindiffusion}) $\sim$ (\ref{eq_spin current}), simulations with various $D$ and $\tau _{\rm s}$ values were performed (see Appendix for details of simulation results) and temporal spin current integrated across the Bi layer, namely, $\int_{0}^{d_{\rm{Bi}}} J_{\rm{s}}(z,t) dz$, was obtained.
 
Figs. 5(a) and 5(b) show the simulated temporal spin current induced by the demagnetization of the Co layer and photo-spin injection into the Bi layer, respectively. 
Figs. 5(c) and 5(d) show the Fourier transformation spectra for spin currents corresponding to Figs. 5(a) and 5(b), respectively. 
The solid curve represents the result fitted with the Lorentzian function to evaluate the bandwidth $\Delta f$. 
Obtained peak values and bandwidth of the spin current plotted as a function of Bi thickness are shown in Figs. 5(e) and 5(f). 
Simulation results roughly agree with experimentally observed Bi thickness dependence. 

\subsection{Discussion}

The initial sharp increase in the demagnetization-induced spin current is due to the spin diffusion in the Bi layer. 
The decrease over 20 nm was mainly caused by the decline in the spin current generated by the demagnetization of the Co layer as mentioned above [depicted by the solid blue symbols in Figs. 4(e) and 5(e)]. 
The photo-spin conversion-induced spin current in the thin region is negligible, which is attributed to the small light absorption in the Bi layer (see Appendix 3 for the laser absorption profile). 
In the thick region, the peak value of the spin current induced by photo-spin conversion gradually increases, which corresponds to the spin diffusion in the Bi layer [open red symbols in Figs. 4(e) and 5(e)]. 
Since spin-current reaches maximum when $d_{\rm Bi}$ is around 80 - 100 nm, which is quite thick compared with light penetration depth of Bi $\sim $ 16 nm, the increasing trends in Figs. 4(e) and 5(e) are not due to light peneratration effect.
Similarly, $\Delta f$ remains almost constant for the spin current induced by demagnetization and decreases with an increase in $d_{\rm Bi}$ for the spin current induced by photo-spin injection, which is obtained in the simulation [Figs. 4(f) and 5(f)].
The parameters used in Figs. 5(e) and 5(f) are $D = 1.2 \times 10^{-3}$ m$^{2}$/s and $\tau_{\rm{s}} = 0.04$ ps for the demagnetization-induced spin current and $D = 1.2 \times 10^{-3}$ m$^{2}$/s and $\tau_{\rm{s}} = 4$ ps for the photo-spin conversion-induced spin current. 
$D$ used in the simulation is obtained based on the electrical conductivity of the sample (see Appendix 2) and the Wiedemann-Franz law.
$\tau_{\rm{s}}$ values were not evaluated by fitting calculation, but extracted by comparing the $d_{\rm Bi}$ dependences of spin-current intensity and $\Delta f$ obtained in the experiment and calculation with changing parameter values an order of magnitude different (see Appendix 5).
\color{black}Extracted quantities such as $\tau _{\rm s}$ mentioned above are not rigorous and have likely an uncertainty due to limitation of modeling.
Backflow spin current observed in photo-spin conversion induced spin-current [opposite polarity at around $\sim$ 1 ps in Fig. 4(b)] cannot be reproduced by the spin-diffusion modeling.
Even with the above-mentioned uncertainty, $d_{\rm Bi}$ dependence of spin current intensity and bandwidth can be explained qualitatively by the spin-diffusion simulation as shown in Fig. 5(e) and 5(f). \color{black}
Note that the $\tau_{\rm{s}}$ values used in the two experiments to explain the experimental results are two orders of magnitude different from each other. 
It should be also mentioned that possible explanation of obvious change in THz waveform by increasing Bi thickness in Fig. 2(d) is attributed to large $\tau _{\rm s}$ of electron spins excited by photo-spin conversion in Bi.
In fact, the increasing slope of the peak spin current value for demagnetization is much higher than that for photo-spin conversion, which is due to different spin relaxation lengths in the two experiments. 
This indicates that the spin relaxation length (here we discuss $\sqrt{D\tau _{\rm s}}$) for demagnetization-induced spin current is one order of magnitude shorter than that for photo-spin injection-induced spin current.

%We discuss the possible origin of the different spin relaxation lengths in the two experiments. 
The differences in the spin relaxation property should be related to the energy level of the spin-transport for optically excited electron spins. 
The electron spin characteristics at the Fermi level are different from those at the optically excited state. 
The mechanism behind the spin current generated by laser-induced demagnetization is an $s-d$ exchange coupling. 
Optically-excited electron relaxes back to the Fermi-Dirac distribution in several tens $\sim $ hundreds of fs\cite{Mueller_2013}.
Then, the loss of angular momentum in local magnetic moment due to demagnetization transfers its angular momentum to mobile $s$-electrons owing to the $s-d$ exchange coupling\cite{Beens_2020}.
Although a shorter spin relaxation length of the laser-induced spin current in ferromagnet/nonmagnet heterostructures, which might be attributed to optically excited electrons, has been observed\cite{Li_2018,Schellekens_2014, Gorchon_2022}, 
the energy level of the electron spins may be close to the Fermi level.
On the other hand, spin current generated by photo-spin conversion is possibly carried by optically excited electron spins in Bi. 
The recombination time of optically excited electrons with 800 nm wavelength light has been $\sim $ 4 ps\cite{Timrov2012}, similar time scale of spin relaxation time used in this study, indicating that spin relaxation is mediated by recombination of optically excited electrons and holes.
%The recombination of electron-hole is $\sim$ 4 ps\cite{Timrov_2012}. The spin relaxation lengths in the laser excited experiment were relatively long. 
%In general, the relaxation time of excited carriers is shorter than that at the Fermi level\cite{Zhukov_2006}. The spin relaxation length of Cu, evaluated using TRMOKE and THz measurements, was 1.3 $\sim$ 50 nm\cite{Li_2018,Schellekens_2014, Gorchon_2022}. This value is much shorter than $\textgreater$ 100 nm obtained from the transport measurement\cite{Bass_2007}. 
The abovementioned fact indicates that spin relaxation length of optically excited spins in Bi is likely longer than those near the Fermi level, which possibly stems from the semi-metallic characteristics of Bi. 

\section{Conclusion}

In this study, two kinds of THz emissions from Bi/Co bilayer films induced by the spin current generation due to demagnetization and photo-spin conversion effect with various Bi thicknesses were investigated simultaneously. 
%Different Bi thickness dependences of peak intensity and bandwidth for spin current were found across two experiments. 
The spin current peak intensity and bandwidth were discussed based on the spin-diffusion simulation with different spin current sources, namely, the demagnetization of Co and photo-spin conversion in Bi. 
It is revealed by the experimental and simulation results that the spin relaxation length of electron spins excited by the photo-spin conversion in Bi is much longer than that induced by the demagnetization of Co, which might be attributed to the semi-metallic characteristics of Bi.

\begin{acknowledgments}
This study is partially supported by KAKENHI (19K15430, 21H05000) and X-NICS of MEXT JPJ011438. K. I. acknowledges Grant-in-Aid for JSPS Fellow (22J22178) and GP-Spin at Tohoku University, S. I. acknowledges the Murata Science Foundation, FRIS Creative Interdisciplinary Collaboration Program in Tohoku University, and JST, PRESTO Grand Number JPMJPR22B2. S. M. acknowledges CSRN in CSIS at Tohoku University.
\end{acknowledgments}

\appendix

\section*{Appendix}
\subsection{Magnetic property}
The magnetic property was evaluated with VSM measurements. Figure 6(a) illustrates the magnetic hysteresis loops for Bi($d_{\rm{Bi}}$)/Co(5) samples with varying Bi thicknesses. The saturation magnetization $M_{\rm{s}}$ was evaluated from the magnetic hysteresis loops and plotted as a function of Bi thickness in Fig. 6(b). The shape of magnetic hysteresis loops and value of saturation magnetization remained approximately constant with respect to Bi thickness. The average saturation magnetization was 1.2 MA/m.

\begin{figure}[ht]
\begin{center}
\includegraphics[width=7cm,keepaspectratio,clip]{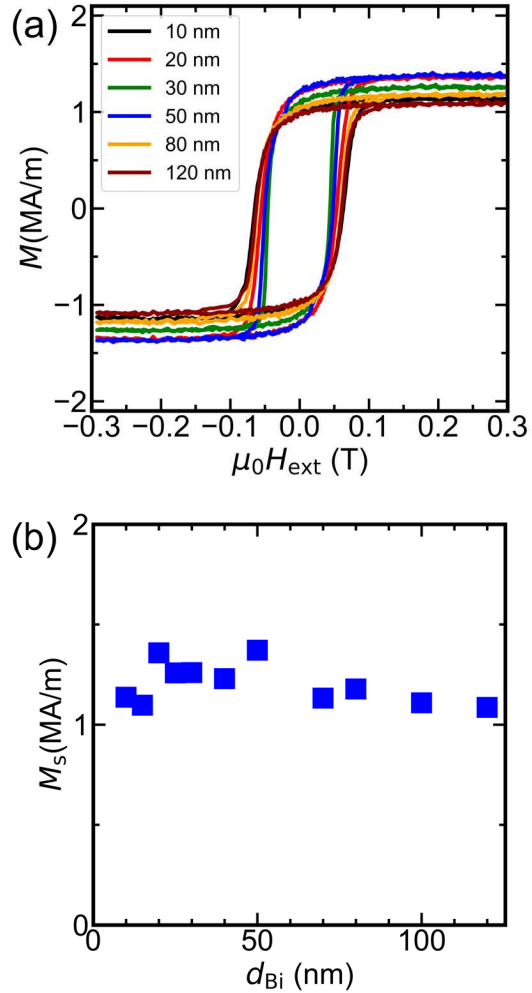}
\end{center}
\caption{(a)In-plane magnetic hysteresis loops of Bi($d_{\rm{Bi}}$)/Co(5)/MgO(2)/Ta(2) (in nm) with various Bi thickness $d_{\rm{Bi}}$. (b)The saturation magnetization obtained from the magnetic hysteresis loops plotted as a function of $d_{\rm{Bi}}$.}
\label{f6}
\end{figure}

\subsection{Electrical conductivity of the sample}

The electrical conductivities of thin film samples were evaluated using the four-point probe method. 
Figure 7 plots the electrical sheet conductivity as a function of Bi thickness $d_{\rm{Bi}}$. 
The slope of the sheet conductivity changes at approximately $d_{\rm Bi}$ = 30 nm.
These data were used to calculate impedance $Z$ [Eq. (\ref{eq_Z})].
The electrical conductivity at thicker regions was evaluated to be $1.9 \times 10^4 $ $\Omega ^{-1} \cdot \rm{m}^{-1}$ based on the slope.

\begin{figure}[ht]
\begin{center}
\includegraphics[width=7cm,keepaspectratio,clip]{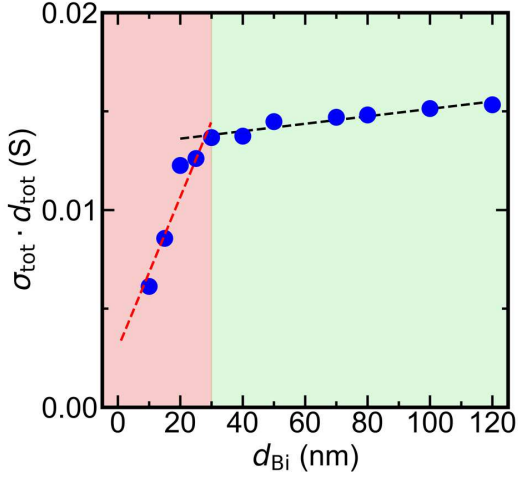}
\end{center}
\caption{Electrical sheet conductivities of thin films plotted as a function of $d_{\rm{Bi}}$. The red and black dashed lines represent linear fitting in thin region and thick region, respectively.}
\label{f7}
\end{figure}

\subsection{Refractive index and absorption of light}

\begin{figure}[ht]
\begin{center}
\includegraphics[width=7cm,keepaspectratio,clip]{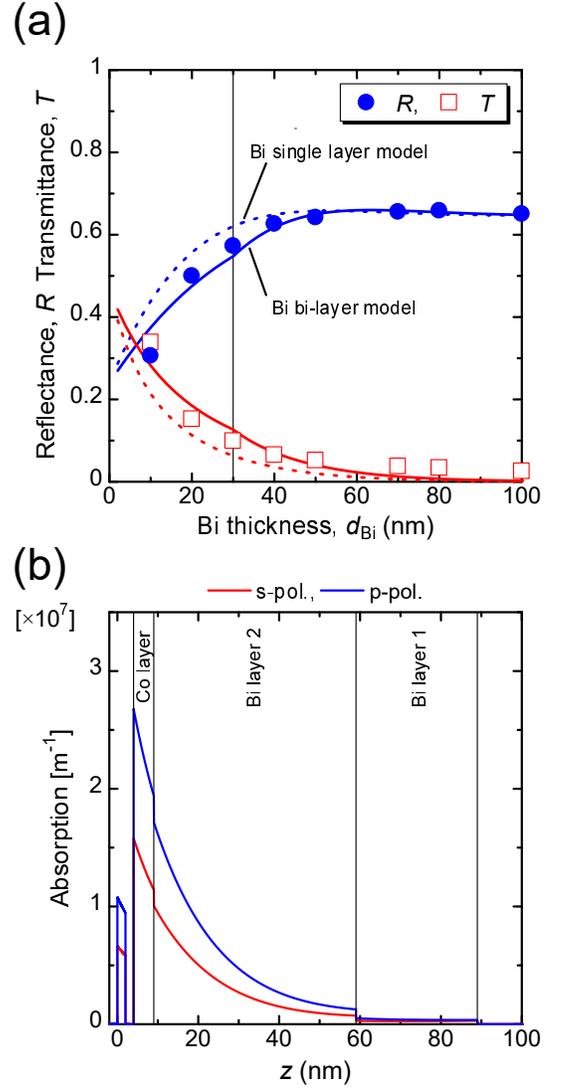}
\end{center}
\caption{ (a)Bi thickness dependence of reflectance and transmittance. The dashed and solid curves denote calculated results using transfer matrix method\cite{Zak_1990} for Bi single layer model and Bi bilayer model, respectively. (b)Light absorption profile calculated using Bi bilayer model for 80-nm-thick Bi sample. }
\label{f8}
\end{figure}

Fig. 8(a) shows the experimentally obtained reflectance [blue solid circles] and transmittance [red open circles] plotted as functions of Bi thickness $d_{\rm{Bi}}$. 
The dashed curves denote $R$ and $T$ values calculated using the transfer matrix method\cite{Zak_1990} with refractive indices listed in Table 1 for Glass sub./Bi2 ($d_{\rm Bi}$)/Co/MgO/Ta thin film (Bi single layer model). 
The calculated results were consistent with those obtained experimentally in the thick region; however, there was a slight discrepancy in the thin region at $d_{\rm Bi}$ $<$ 30 nm. 
A similar trend was observed for electrical conductivity: it varied at around $d_{\rm Bi}$ = 30 nm (See Appendix 2). 
To explain the discrepancy in the thin region, the Bi layer was divided into interfacial ($\textless$ 30 nm) and bulk layers ($\textgreater$ 30 nm). 
The refractive index of the bulk Bi layer (Bi2) was taken from literature while that of the interface Bi layer (Bi1) was obtained by fitting the experimental $d_{\rm Bi}$ dependence of $R$ and $T$.
The solid curves depicted in Fig. 8(a) correspond to $R$ and $T$ values calculated using the Bi bilayer model. 
$R$ and $T$ values at the thin region were explained well using the Bi bilayer model compared with the explanation provided by the Bi single layer model.
Fig. 8(b) shows the calculated light absorption profile $a(z)$ for the Bi(80)/Co(5) sample. 
The light absorption of circularly polarized light corresponds to an average of light absorption with s- and p-polarizations, which is used to calculate the photo-spin injection [Eq. (\ref{eq_Q_p})]. 

\begin{threeparttable}[hb]
  \caption{Refractive index (800 nm) used for transfer matrix calculation}
  \begin{tabular}{cccc}
    \hline\hline
    Material & \hspace{5mm}Thickness (nm)\hspace{5mm} & Refractive index   \\
    \hline \hline
    Ta & 2  & 1.11 + $i$ 3.48\cite{OC_Ta} \\
    MgO & 2 & 1.73  \cite{OC_MgO}\\
    Co & 5 & 2.49 + $i$4.80 \cite{OC_Co} \\
    Bi2 & $\Biggl\{ $ \begin{tabular}{lll} $d_{\rm Bi}$ -30 & $\cdots $ & $d_{\rm Bi}$ $>$ 30 \\ 0 & $\cdots $ & $d_{\rm Bi}$ $\leq $ 30   \end{tabular} & 2.78 + $i$3.79 \cite{OC_Bi} \\
    Bi1& $\Biggl\{ $ \begin{tabular}{lll} 30 & $\cdots $ & $d_{\rm Bi}$ $>$ 30 \\ $d_{\rm Bi}$ & $\cdots $ & $d_{\rm Bi}$ $\leq $ 30 \end{tabular} & 1.9 + $i$2.2  \\ 
    Glass sub. & 500000 & 1.5 \\
    \hline\hline
  \end{tabular}
  \begin{tablenotes}
  \item{*} Real and imaginary part of refractive index at interface Bi layer 1 was obtained by fitting to experimental reflectance and transmittance data.
  \end{tablenotes}
\end{threeparttable}

\subsection{Ultrafast demagnetization via TRMOKE measurement}
Laser-excited ultrafast demagnetization of Co was evaluated using the TRMOKE measurement. Fig. 9(a) shows normalized magnetization dynamics for the Bi(15)/Co(5) bilayer film. The red solid curve denotes the fitting result obtained with the equation\cite{Longa_2007,Iihama_2015}
\begin{align}
&\frac{\Delta \theta _{\rm K}(t)}{\theta _{\rm K}} = \biggl[ \biggl\{ \frac{\Delta m_{1}}{\sqrt{1+t/\tau _0}}-\frac{\Delta m_{2}\tau_{\rm{E}}-\Delta m_1\tau_{\rm{M}}}{\tau_{\rm{E}}-\tau_{\rm{M}}}e^{-t/\tau_{\rm{M}}} \notag \\
&\hspace{1cm} -\frac{\tau_{\rm{E}}(\Delta m_{1}-\Delta m_{2})}{\tau_{\rm{E}}-\tau_{\rm{M}}}e^{-t/\tau_{\rm{E}}} \biggr\} \Theta(t) +\Delta m_{3}\delta(t) \biggr] \star G(t),
\label{eq_demag}
\end{align}
where $\Delta m_{1}, \Delta m_{2}$, and $\Delta m_{3}$ represent the change in the magnetization at equilibrium state, maximum demagnetization, and state spin filling effect, respectively. $\tau_{\rm{M}}$ and $\tau_{\rm{E}}$ denote the demagnetization and remagnetization times, respectively. $\Theta(t), \delta(t)$, and $\star$ are the Heaviside step function, Dirac delta function, and convolution product, respectively. The Bi thickness dependence of demagnetization amplitude is well described with inverse total thickness, i.e., $\propto (d_{\rm{Bi}}+d_{\rm{Co}})^{-1}$ as shown in Fig. 9(b). The red solid curve denotes the fitting result.

\begin{figure}[ht]
\begin{center}
\includegraphics[width=7cm,keepaspectratio,clip]{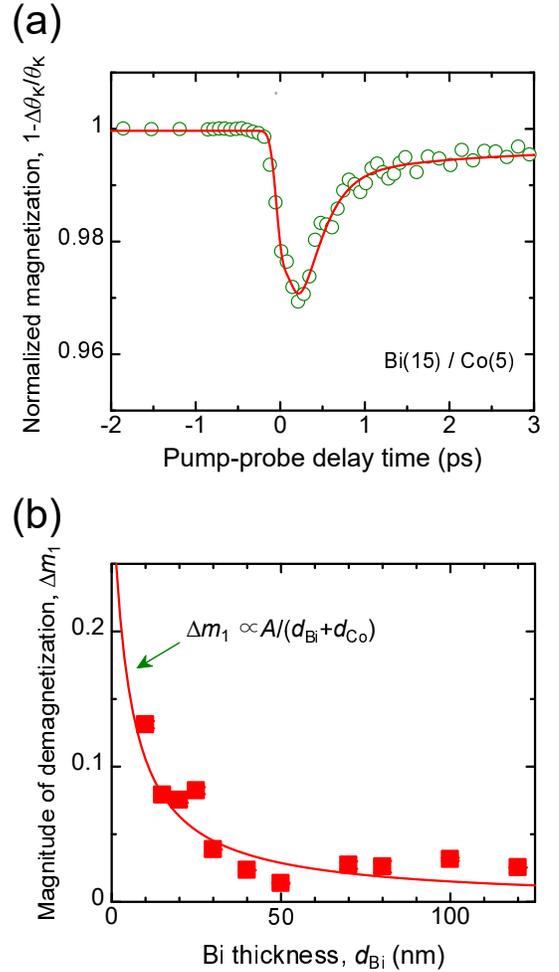}
\end{center}
\caption{ (a)Demagnetization dynamics for Bi(15)Co(5) bilayer film. The solid curve was fitting result with eq. (\ref{eq_demag}). (b)Magnitude of magnetization plotted as a function of Bi thickness. The solid curve represents a fitted result using a function inverse proportional to total thickness of Co and Bi.}
\label{f9}
\end{figure}

\subsection{Simulation with various parameters}

We performed a simulation with various diffusion constant $D$ and spin relaxation time $\tau_{\rm{s}}$ values. 
$D$ was assigned the values of 1.2 $\times$ 10$^{-4}$, 1.2 $\times$ 10$^{-3}$, and 1.2 $\times$ 10$^{-2}$ m$^2$/s. 
$\tau_{\rm{s}}$ was assigned the values of 4, 40, 400 fs, and 4 ps.
Figs. 10(a) and 10(b) show the peak value and bandwidth of spin current induced by the demagnetization of the Co layer plotted as a function of Bi thickness $d_{\rm{Bi}}$ with different $D$ and $\tau_{\rm{s}}$ values. 
Figs. 11(a) and 11(b) show the peak value and bandwidth of spin current generated by photo-spin injection in the Bi layer plotted as a function of $d_{\rm{Bi}}$ with different $D$ and $\tau_{\rm{s}}$ values. 
$D$ can be obtained by the Wiedemann–Franz law given by, $D=L\sigma / \gamma _{\rm e}$, where $L, \sigma$, and $\gamma_{\rm{e}}$ are the Lorenz number, electrical conductivity, and electronic heat capacity, respectively. 
When we use $\sigma $ = 1.9 $\times$ 10$^{4}$ $(\Omega\cdot$m$)^{-1}$ evaluated by the four-point probe measurement and $\gamma _{\rm e}$ = 0.37 $\rm{J}\cdot \rm{m}^{-3}\cdot \rm{K}^{-2}$ for Bi taken from the literature\cite{Kittel}, $D$ = 1.2 $\times $ 10$^{-3}$ $\rm{m}^2 / \rm{s}$ is obtained. 

\begin{figure*}[ht]
\begin{center}
\includegraphics[width=0.95\textwidth,keepaspectratio,clip]{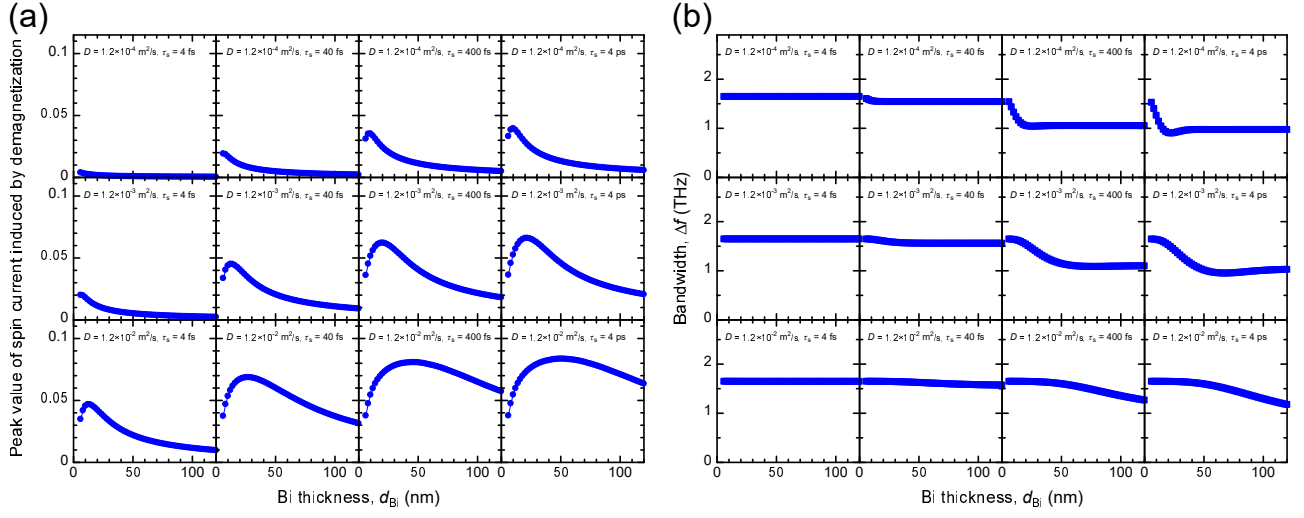}
\end{center}
\caption{ Spin current (a) peak intensity and (b) bandwidth induced by laser-induced demagnetization plotted as a function of Bi thickness $d_{\rm Bi}$ with different diffusion constant $D$ and spin relaxation time $\tau _{\rm s}$.}
\label{f10}
\end{figure*}

\begin{figure*}[ht]
\begin{center}
\includegraphics[width=0.95\textwidth,keepaspectratio,clip]{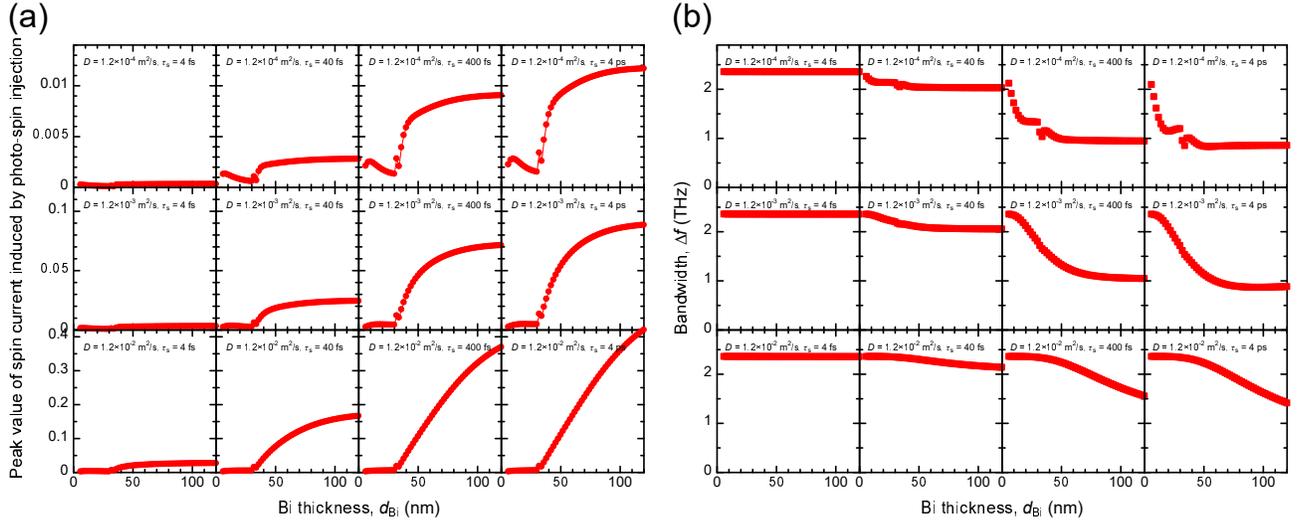}
\end{center}
\caption{ Spin current (a) peak intensity and (b) bandwidth induced by photo-spin injection into Bi plotted as a function of Bi thickness $d_{\rm Bi}$ with different diffusion constant $D$ and spin relaxation time $\tau _{\rm s}$.}
\label{f11}
\end{figure*}

\end{document}